# Analysis of Link Formation, Persistence and Dissolution in NetSense Data

Ashwin Bahulkar[1] and Boleslaw K. Szymanski[1,2]
[1]Department of Computer Science & NEST Center
Rensselaer Polytechnic Institute
Troy, NY, USA
[2]Społeczna Akademia Nauk
Łódź, Poland
{bahula,szymab}@rpi.edu

Omar Lizardo[3,5], Yuxiao Dong[4], Yang Yang[4], and Nitesh V. Chawla[4,5]
[3]Department of Sociology & iCeNSA
[4]Computer Science & Engineering Departments & iCeNSA
University of Notre Dame, Notre Dame, IN, USA
[5]Wrocław University of Technology, Wrocław, Poland
{olizardo,ydong1,yyang1,nchawla}@nd.edu

*Abstract*— We study a unique behavioral network data set (based on periodic surveys and on electronic logs of dyadic contact via smartphones) collected at the University of Notre Dame. The participants are a sample of members of the entering class of freshmen in the fall of 2011 whose opinions on a wide variety of political and social issues and activities on campus were regularly recorded—at the beginning and end of each semester—for the first three years of their residence on campus. We create a communication activity network implied by call and text data, and a friendship network based on surveys. Both networks are limited to students participating in the NetSense surveys. We aim at finding student traits and activities on which agreements correlate well with formation and persistence of links while disagreements are highly correlated with non-existence or dissolution of links in the two social networks that we created. Using statistical analysis and machine learning, we observe several traits and activities displaying such correlations, thus being of potential use to predict social network evolution.

## I. INTRODUCTION

Renewed attention to dynamics in social network analysis and network science in the recent literature has brought back a concern with classical questions regarding the origins of personal relationships [2, 7], as well as the factors that account for their temporal persistence [8, 9, 11]. This has re-opened central issues under-emphasized in classical network theory: the problem of the emergence of network ties, the problem of the evolution of social relationships over time and the factors that contribute to dynamic tie persistence and decay.

Standard contagion-based models propose that persons become more similar because they share a social tie [14]. From this perspective networks evolve to behavioral commonality via influence-based processes. The key assumption is that the network itself evolves independently of behavior and attitudes. In contrast to this assumption, network co-evolution models suggest that in the very same way in which network ties may result in the strengthening or weakening of behavioral propensities, it is also possible that previously existing agreement on certain behavioral propensities will be responsible for the strengthening or weakening of network ties. Social relationships have effects on behavior and attitudes, but behavior and attitudes may also have an effect on the structure of social relationships.

A key problem in empirically investigating this side of the co-evolution process is that agreement (or disagreement) on behavioral propensities may affect social ties via two distinct mechanisms. First, previous agreement may generate *new* ties were none previously existed. This process has been referred to as (status, value, or choice) "homophily" in classical [2] and contemporary social network theory [11, 12]. Here, preexisting behavioral and attitudinal commonalities facilitate the tie formation process. Second, as has been noted by other analysts agreements may also affect network dynamics via a pruning or negative selection process, whereby existing disagreements either prevent new ties from forming or lead to faster decay of low agreement ties in relation to high agreement ties [9].

With data taken at a single point in time, it is impossible to tease apart whether network evolution is driven by homophily or selective decay. This leads to threats to causal inference or to over-inflated estimates of influence and contagion processes [9, 13]. In this study, we leverage unique data containing over-time information on opinion, attitudinal, and behavioral agreement, as well as unobtrusively collected data on electronic communication to establish whether network dynamics are more deeply affected by value homophily, unfriending dynamics or both at the same time.

## II. NETSENSE DATA

The NetSense data that used in this study consists of students reporting their on-campus activities, personal traits and interests, as well as views and opinions on various social issues at the beginning of every school semester from the fall 2011 to the spring 2013. The data also contains a record of the calls made and texts messages sent between students participating in the study during the same period.

From these data, we create two social networks. The first one is a student communication activity network that contains a link connecting two students if calls made and text messages exchanged between them over the given semester exceed a modest threshold. The second network is a friendship network based on traditional name-generator surveys in which each student in each semester lists friends on and off campus. These two social networks are updated for each semester covered by the data, thus creating four snapshots for each of the two evolving networks.

Our objective is to examine how strongly the agreements and disagreements between pairs of students on their on-campus activities, personal traits and their views are correlated with the formation, persistence and dissolution of links in the two social networks derived from the NetSense data.

We used statistical analysis and machine learning techniques to find which network structures and agreements over which traits are correlated with formation and deletion of links in the two networks.

We divide our analysis into two sub-problems:

1. *Link Prediction*: Predicting if a link would form in the future. Section III discusses link prediction [4].

2. *Prediction of Link Persistence*: Predicting if an existing link would persist or get dissolved in the future. Section IV discusses link persistence prediction [6].

### A. Call and Text Messaging Data

We used the NetSense records calls made and text messages exchanged by students from August 2011 to May 2013. Each record consists of an entry for each call or text message recorded, specifying the date of the communication, the sender and receiver of the communication and the duration or length of the communication.

### B. Data on Node Attributes

Students participating in the Netsense study filled out a survey at the beginning of each semester. Survey questions were about the students' family background, major pursued in Notre Dame University, classes taken and activities on campus, as well as their views on several social issues, their political inclination, and their behavioral traits.

For each student we selected the following attributes from the NetSense data:

*Student background*:
- Major in the Notre Dame programs
- Behavioral traits: Is-Talkative, Is–Outgoing, Is-enthusiastic
- Family income
- Race
- Religion followed

*Social and political views on:*
- Politics
- Abortion
- Marijuana legalization
- Homosexuality and gay marriage
- welfare and social security
- Racial equality and affirmative action

*Habits and Lifestyle:*
- Drinking habits
- Time spent weekly on activities like studying, partying, socializing, volunteering, campaigning for social causes and exercising
- Classes taken and clubs joined

### C. Student Communication Activity Network

We created an evolving student communication activity network with snapshots taken in each semester. An edge exists between a pair of students if a sufficient number of calls or text messages are made between them during the particular semester

We created network snapshots for four semesters: Fall 2011 semester ranging from August 2011 to December 2011, Spring 2012 semester lasting from January 2012 to May 2012, Fall 2012 semesters ranging from August 2012 to December 2012, and Spring 2013 semester lasting from January 2013 to May 2013. Since very few calls were made during the summer of 2012, we did not create a network for the summer semester. Table I shows the sizes of the snapshots of this evolving network[1].

TABLE I. SIZE OF NETWORK SNAPSHOTS IN EACH SEMESTER

| Network Snapshot | No. of Nodes | No. of Edges |
|---|---|---|
| Fall 2011 | 205 | 346 |
| Spring 2012 | 193 | 311 |
| Fall 2012 | 189 | 207 |
| Spring 2013 | 169 | 167 |

### D. Converting Node Attributes to Edge Attributes

To measure how attributes correlate with formation or deletion of links, we examine if similar nodes form links and dissimilar nodes dissolve links. To do this, we need to measure the agreement levels between nodes, which form edges, over all the traits listed in Section IIB.

The agreement values are normalized between 0 and 1, where agreement on the binary traits is denoted by 1, while disagreement is denoted as 0. For non-binary attributes, 0 denotes the most disagreeing values and 1 stands for the most agreeing values, while the intermediate values are selected from the range between these two extremes.

### III. LINK PREDICTION ON NETSENSE

We want to find out which attributes correlate strongly with formation of links. To this end, we classify existing and non-existing edges into categories depending on their status in the current and future semesters and compute traits agreements of each category of edges. Finally, we observe

---

[1] We did similar analysis on the friendship network as well, observations and results for which were very similar to those of the communication network.

which trait agreements are best correlated with the process of link formation.

*A. Observations*

We investigate whether there are differences in the average numbers of attribute agreements between endpoints of existing, to-be-formed, and non-existing edges, where to-be-formed edges are those, which do not exist in a currently considered semester, but are created in the succeeding one. We compute both the total number of traits over which endpoints of an edge agree and the number of neighbors common to endpoints of an edge and then average theses values for the three classes of edges defined above.

We compare the total number of agreements across dyads and find that existing edges agree the most, on average, on the relevant traits. This is in accordance with the principle of the homophily mechanism outlined earlier [2, 11]. We further observe that to-be-formed edges have much higher level of agreement on traits than edges that do not form in the future. This suggests that homophily plays a role in the network evolution process, making some ties more likely to form than others [7, 11]. This also indicates that we could use edge agreements to predict the formation of links as these edges could easily be differentiated from edges that do not form. We also observe that existing edges have a lot more friends in common as compared to non-existing edges, and to-be-formed edges have a lot more friends in common as compared to edges, which do not form.

In Figures 1 and 2, we plot the average number of edge agreements and the average number of common neighbors for existing edges, non–existing but to-be-formed edges and non-existing edges for each semester. The NetSense data contains four semesters, but we have predictions only for semester 1 to 3, since we need one "future" semester to distinguish between non-existing but to-be-formed edges and the edge that are truly non-existing. The detailed statistics for each trait are provided in the Appendix.

*B. Link Prediction Using Machine Learning*

Since we want to know which attributes correlate with the formation of links, we need to classify links based on the attribute agreements and then observe which attributes were more important during classification.

Observations from section IIIA imply that it is certainly possible to predict formation of edges based on the number of agreements over all traits. We used different classifiers to predict formation of edges. The steps that we took to reach results are listed below.

Task: Predict whether an edge would form given the agreement between edges over the attributes described above, number of agreements and number of common neighbors, which are the features for classification. There are 29 features in total. They include all agreements over the 27 attributes listed earlier, plus the agreements over the number of common neighbors and the total number of agreeing attributes.

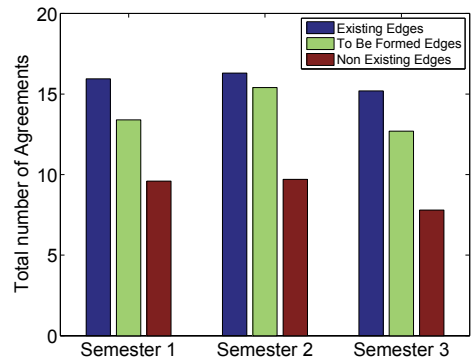

Fig. 1. The semester-wise average number of agreements for existing, to-be-formed and non-existing edges.

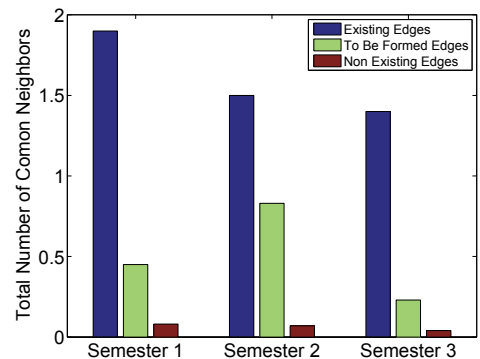

Fig. 2. The semester-wise average number of common neigbors for endpoints of existing, to-be-formed and non-existing edges.

*Dataset specifications*: For semester 1 to 3, we use links, which formed in the succeeding semester as positive examples, and links, which did not form in the succeeding semester as negative examples of edge formation. We split the data putting 80% of data as a training dataset and the remaining 20% of the data as a test dataset where the remaining 80 % was used as training data. Counting all the possible edges, we found in the training dataset about 55,000 negative examples but only about 250 positive examples.

*Imbalanced data*: The data set that we obtained is severely imbalanced, with too many negative examples. We restrict the number of negative examples by considering only those edges that lie in a certain neighborhood of a node. For example, limiting the potential edges whose endpoints are separated by the distance of at most three hops in the graph reduces the number of negative examples to about 10,000. Using only those edges whose endpoints are separated by a distance of at most two hops, we reduce the number of negative examples further to 2,500. We perform experiments on the datasets obtained by using edges whose endpoints are separated by at most two hops [1].

*Need for Dimensionality Reduction*: Using the 29 agreement based features yields an accuracy close to 99%, while giving a very poor recall of 0-20 %. This indicates that none of the features are good indicators of links by themselves. Using the features all by themselves, the classifiers could not predict any links. However, when we use dimensionality reduction by Singular Value Decomposition (SVD) [16], we are able to achieve good recall, which however comes with an acceptable drop in accuracy. This indicates that a combination of several features is needed to achieve good correlation with the creation of links. We feed the features in the new feature space to the classifier.

*Classification techniques used*: We use logistic regression, linear and kernel SVM, K-NN, random forests, naïve bayes classifier and ensemble method based classifiers [2] using the dimensions in the new space mapped by SVD.

### C. Results of Machine Learning

We observe that SVM and logistic regression work the best when we use a high number of Eigen vectors (see the value marked in bold italics in Table II). With low number of Eigen vectors, we are able to get good accuracies, which however come with a recall of 0. Recognizing links is important for us. Using other classification techniques, we are able to get good accuracies, which however are accompanied by low recall values. Table II shows these results confirming that logistic regression with top 28 Eigen features yields the best results. We also obtained results for the friendship network, and we observe that the results are very similar.

TABLE II. RESULTS FOR LINK PREDICTION

| Classification Method | No SVD | | Top 2 Eigen Features | | Top 28 Eigen Features | |
|---|---|---|---|---|---|---|
| | *Accuracy* | *Recall* | *Accuracy* | *Recall* | *Accuracy* | *Recall* |
| SVM | 98.0 | 0.0 | 98.0 | 0.0 | *82.9* | *72.0* |
| Logistic regression | 98.0 | 0.0 | 98.0 | 0,0 | *76.7* | *75.9* |
| k-NN | 97.0 | 11.6 | 96.8 | 16.4 | 97.0 | 11.6 |
| Random forests | 98.0 | 6.9 | 97.9 | 18.6 | 98.0 | 9.3 |
| Naïve-Bayes | 98.0 | 0.0 | **95.3** | **55.2** | 90.2 | 48.8 |
| SVM- RBF Kernel | 98.0 | 0,0 | 98.0 | 0,0 | 98.0 | 0.0 |
| Ensemble of Classifiers | 98.1 | 9.3 | 98.1 | 9.3 | 98.1 | 9.3 |

### D. Ranking of Features

The equation for Singular value Decomposition (SVD) is as follows:

$$A = U S V^T$$

$U$ contains the left singular vectors which are the features in the new feature space. $S$ stores the singular values, $V$ contains the right singular vectors. The right singular vectors exist for each of the features in the new space express the new features in terms of the original features. We also have the weights returned by the classifier. To obtain the importance of each original feature, we square each value in the right singular vectors and multiply the right singular vectors by the weight vector to obtain weights for each of our original features.

The right singular vectors are contained in the matrix $V$. Each column represents a feature in the new space; it contains weights for each of the original features. If we have $m$ original features and $k$ new features, the dimension of this matrix is $m*k$. Moreover, with $k$ new features, the weight vector, $W$, returned by the classifier has $k*1$ dimensions. By performing matrix multiplication, $V^2 \times W$, we get the weights for each of the original $m$ features.

We obtained scores for each of our original features using the above method.

Table III lists the scores for each feature. We observed that the average number of common neighbors and the average number of common traits were the most important features, with the highest scores, followed distantly by time nodes spent on activities, parental income, and views on various issues. Religion, race, major and classes taken together have low scores, so appear to matter only a little for link prediction.

The values in Table III imply that acquaintances and contacts are more likely to form between people who have common friends and have a significant amount of common interests and traits.

When using different numbers of Eigen features for classification, rankings of features are a little bit unstable, especially in case of the smaller ones. Rankings for the first 10 traits barely change, but the others quite often move their rankings a bit.

### IV. PREDICTION OF PERSISTENCE ON NETSENSE

We are interested in knowing the features on which agreements between endpoints of an edge correlate strongly with the persistence of this edge. First, we find whether the edges which after being formed persist for a long time differ in the total number of agreements of features between their endpoints from the edges that after being created dissolve relatively quickly and disappear. Next, we also evaluate impact of single features on persistence using the following task.

*Link Persistence Sensitivity Task*: In this task we ranks features according to their impact on link persistence. In addition, we also measure the differences between edge persistence based on the number of calls made and text messages exchanged for both the communication activity network and the friendship network.

TABLE III. RANKING OF FEATURES FOR LINK PREDICTION

| Feature | Scores |
|---|---|
| Number of Common Traits | 10.1 |
| Number of Common Neighbors | 3.2 |
| Hard Drinking | 0.52 |
| Time Spent on Campaigning | 0.52 |
| Time Spent on Volunteering | 0.5 |
| Parental Income | 0.5 |
| Political Views | 0.47 |
| Time Spent at College Clubs | 0.45 |
| Time Spent on Partying | 0.45 |
| Time Spent on Exercising | 0.45 |
| Time Spent Socializing | 0.43 |
| Views on Equality | 0.41 |
| Views on Marijuana Legalization | 0.41 |
| Views on Abortion | 0.39 |
| Views on Gay Marriage Legalization | 0.32 |
| Views on Homosexuality | 0.24 |
| Views on Pre Marital Sex | 0.18 |
| Behavioral Traits | 0.18 |
| Race | 0.113 |
| Religion | 0.09 |
| Views on Social Welfare | 0.002 |
| Major | 0.0018 |
| Classes Taken | 0.0006 |
| Clubs Joined | 0.0002 |

## A. Observations: Feature Disagreement and Network Structure

We want to find disagreements over which features of the endpoints of an edge is strongly corrrelated with the dissolution of this edge. An edge is persistent in a semester when it is also present in the succeeding semester. An edge is dissolving if it does not survive into the succeeding semester. We measure the differences between persisting and dissolving edges with respect to the number of agreeing features. However, there seems to be very little difference between persistent and dissolving edges when it comes to number of agreeing features, except of the difference in the number of common neighbors. Figure 3 illustrates this observation.

## B. Observations: the levels of call and text comunications and their correlation with edge persistence of student communication activity network edges.

We observe that there is a significant difference in the calling and texting frequency in a semester between endpoints of the edges that persist and those that dissolve. Figure 4 and Figure 5 illustrate the differences in average level of these communications between endpoints of persisting edges and dissolving edges.

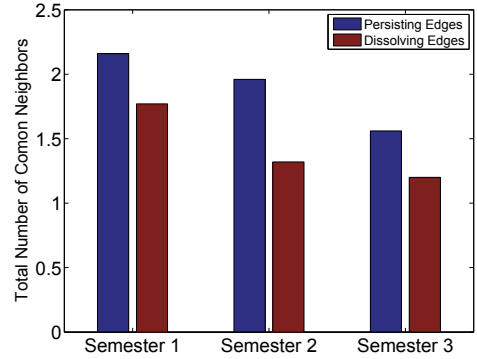

Fig. 3. Student communication activity network: the average number of common neighbors for persistent and disolving edges.

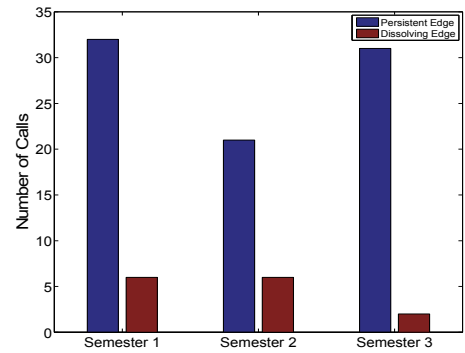

Fig. 4. Student communication activity network: the average number of calls per semester.

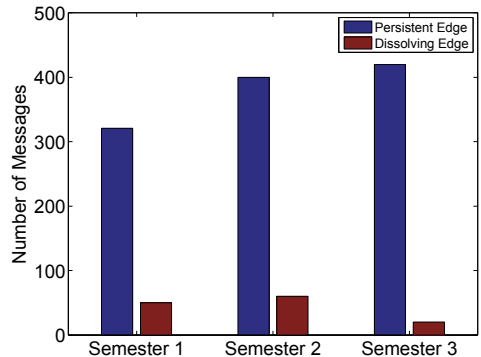

Fig. 5. Student communication activity network: the average number of text messages per semester.

*C. Observations: the levels of call and text comunications and their correlation with persistence of friendship network edges.*

In the student friendship network we observe that friendships, which dissolve, have a significantly lower levels of communications between friends (typically several times lower) than between friends whose friendship edge will persist. Figure 6 and Figure 7 illustrate these differences.

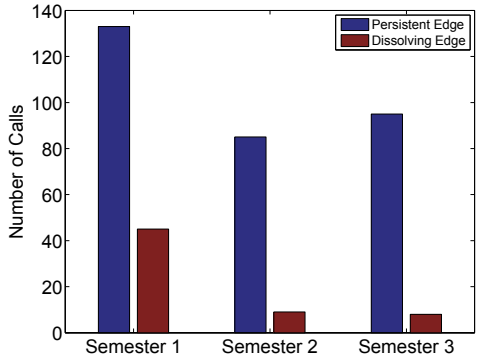

Fig. 6. Friendship network: the average number of calls per semester.

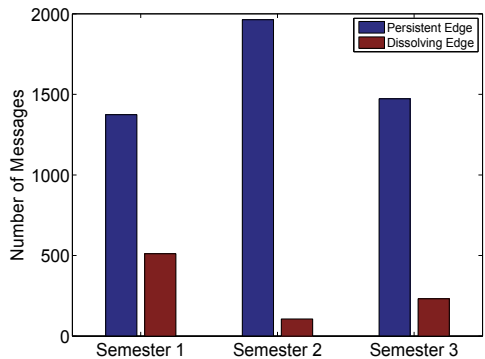

Fig. 7. Friendship network: the average number of text messages per semester.

*D. Machine Learning Techniques for Predicting Edge Persistence*

We use the same machine learning techniques that we applied for link prediction to predict link persistence. The results for classifying persistent and dissolving edges based on differences between features of endpoints of edges is difficult because we find that edge endpoint agreements on a few features could separate these edges. Table 3 shows these results.

The details of this classification task are listed below.

*Task*: Predict whether an edge would persist or not given the feature agreements between its endpoints. The agreement values form the machine learning features for prediction. We have 29 features in total.

*Dataset specifications*: For semesters 1 to 3 we use links, which persisted into the succeeding semester as positive examples, and links, which got dissolved in the succeeding semester as negative examples. We split the data into training set with 80% of the entire dataset, while the remaining 20% of the entire dataset was put aside as test data. The total number of samples in the training dataset was about 800 edges; out of which about 500 were positive examples, while 300 were negative ones.

TABLE IV. RESULTS FOR PREDICTION OF LINK PERSISTENCE

| Classification Method | No SVD | Top 2 Eigen Features | Top 15 Eigen Features | Top 28 Eigen Features |
|---|---|---|---|---|
| SVM | 60.7 | 60.7 | 60.7 | 62.2 |
| Logistic regression | 62.8 | 62.8 | 62.9 | 62.8 |
| k-NN | 59.7 | 60.7 | 57.2 | 58.8 |
| Random forests | 65.2 | 57.2 | 65.2 | 64.7 |
| Naïve-Bayes | 37.2 | 37.2 | 45.7 | 56.2 |
| SVM-RBF Kernel | 58.4 | 67.2 | 56.2 | 62.8 |
| Ensemble of Classifiers | 56.2 | 56.2 | 56.2 | 56.2 |

*Classification techniques used*: We use logistic regression, linear SVM, K-NN, random forests, naïve bayes classifier and ensemble method based classifiers. We do not observe any appreciable differences between results obtained with dimensionality reduction and without it; the classifier accuracies barely change from one case to another. However, SVM using RBF kernel was found to have a slightly higher accuracy when limited Eigen features were used.

*E. Ranking of Features*

To know the importance of each feature, we derive a score for each one as described in section IIID. The top five of the features are listed in Table V.

TABLE V. RANKING OF FEATURES FOR LINK PERSISTENCE

| Feature | Scores |
|---|---|
| The number of Common Traits | 2.86 |
| The number of Common Neighbors | 1.01 |
| Views on Marijuana Legalization | 0.12 |
| Religion | 0.04 |
| Race | 0.03 |

We find that only two features, namely the total number of agreements between edge endpoints on all traits and the number of common neighbors of the edge endpoints are significant for the prediction of link persistence, but even those are having just medium weights. Other features played a much lesser role, as evidenced by very small weights for features ranked from position 3 and below. This is unlike link prediction, where many other features beside the first two were significant as well. In fact the third ranked feature here would not make the top 20 ranked features for link prediction.

## V. CONCLUSION

In this paper, we set out to investigate whether homophily based on pre-existing behavioral and attitudinal agreement or selective pruning based on existing disagreements is the most powerful mechanism in shaping over-time network dynamics. We observed that existing edge endpoint agreement on traits and network structure play an important role in the formation of new links. This is consistent with the claim that the mechanism of value or choice of homophily is a powerful driver of network co-evolution processes [2, 7, 11]. We also found that pre-existing agreements between endpoints traits play an important role in social tie persistence, comparable to and in some cases surpassing the effect of such factors as number of (within-study) common neighbors.

These results do not seem to be driven by the overarching influence of any one attitude, trait, or behavior, but by the overall rate of agreement across a larger number of elements. This is in contrast to sociological models that posit the preponderance of one or two socio-demographic factors, such as race, gender, or religion [12], and more consistent with culture-network co-evolution model which points to the sharing of a large pool of cultural facts as the prime drive of network evolution [15].

Finally, we found a more muted effect of existing disagreement in driving network co-evolution via the pruning or selective decay mechanism. These results indicate that claims to threats to causal inference attributable to this "unfriending" problem, may have been over-stated in the literature [9]. In this respect, our work suggests that analysts should pay more attention to the issue of confusing contagion with homophily processes [13]. Alternatively, it is possible that the homophily mechanism operates at a different (slower) time scale than the pruning process, and that a more temporally fine-grained analysis (e.g., measuring agreement change in weeks rather than months) is required to detect the operation of pruning effects on social ties. This opens up an important avenue for further research on this topic.

## ACKNOWLEDGMENT

We thank Prof. G. Korniss for beneficial discussions. This work was supported in part by the Army Research Laboratory under Cooperative Agreement Number W911NF-09-2-0053 (the Network Science CTA), by the Office of Naval Research (ONR) grant no. N00014-15-1-2640, by the European Commission under the 7th Framework Programme, Agreement Number 316097, and by the Polish National Science Centre, the decision no. DEC-2013/09/B/ST6/02317. The views and conclusions contained in this document are those of the authors.

APPENDIX: *AGREEMENT VALUES FOR FOR ALL PAIRS OF NODES AND FOR EACH ATTRIBUTE.*

For each trait, we compute the average number of agreements between pairs of endpoints of all existing edges, to-be-formed edges and all non-existing edges. We compute these values for the first three semesters, but not for the fourth one for which we do not know which new edges will form and which existing edges will dissolve in the future. Figures 8 to 15 illustrate these statistics for views on drinking habits on marijuna legalization, on gay marriages, on politics, and for the time spent on partying, for the classes taken together, and for the clubs joined together.

For these traits, the differences in agreement between endpoints of edges are large enough to be of value in predicting edge persistence for existing edges and edge formation for currently non-exiting ones.

Figures 8-10 show the agreements that are sufficient for differentiating between existing and non-existing edges. It is interesting to notice that the similarity between existing and to be formed edges increases as the students advance in from semester to semester. By the third semester all three types of edges have the same high level of agreement.

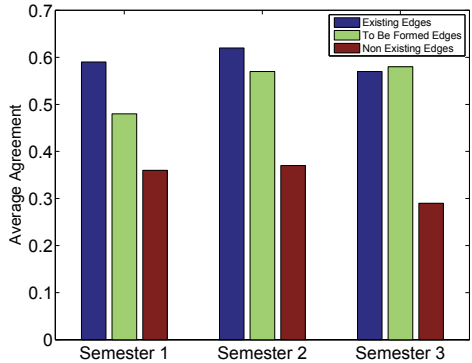

Fig. 8. The agreement between edge endpoints on views on abortion.

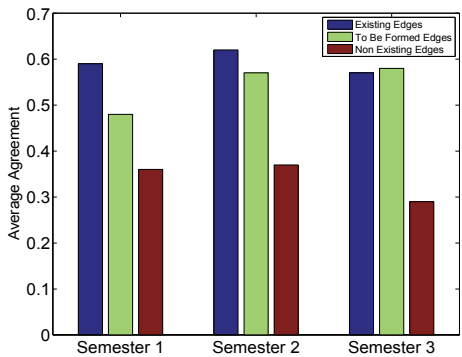

Fig. 9. The agreement between edge endpoints on drinking habits.

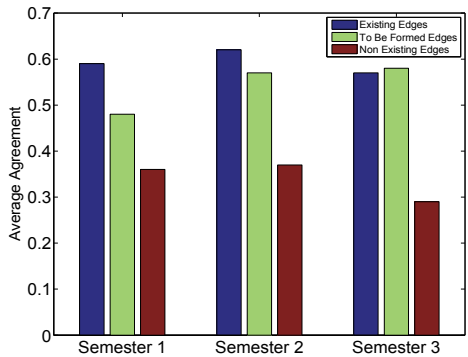

Fig. 10. The agreement between edge endpoints on views on marijuana legalization.

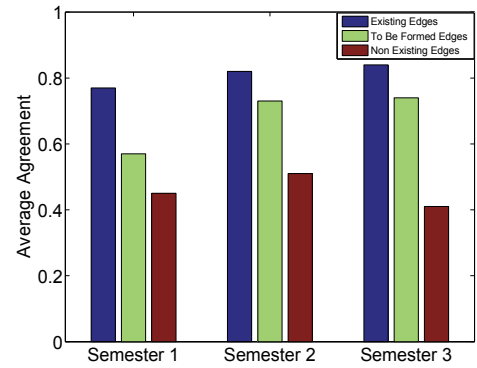

Fig. 11. The agreement between edge endpoints on views on gay marriage.

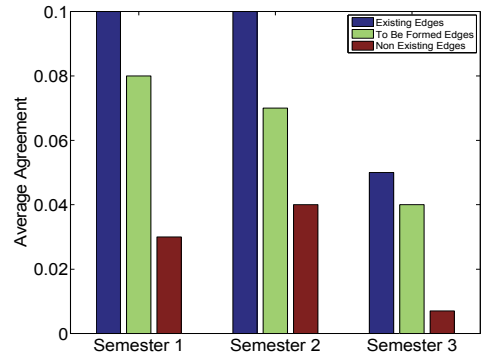

Fig. 12. The agreement between edge endpoints on classes taken together.

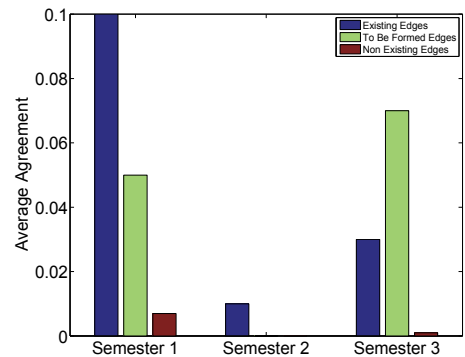

Fig. 13. The agreement between edge endpoints on clubs joined together.

Figure 11 shows the highest level of agreement on traits for views on gay marriage with small differences between existing and to be form edges and larger differences between existing and non-existing edges. In contrast, Figures 12-13 show lowest level of agreement for classes and clubs, because unlike for other traits, there are many options in these to cases. Some agreement on classes and clubs is important for edge formation, but less important for persistence in the later years.